\pdfoutput=1
\documentclass{ws-ijmpc}
\usepackage{mathptmx}
\usepackage{amsmath}
\usepackage{amssymb}
\usepackage{graphicx}
\usepackage{hyperref}
\usepackage{dcolumn}
\usepackage{bm}
\usepackage{color}

\begin{document}

\markboth{Shigang Qiu \& Tao Jia}
{Quantifying the noise in bursty gene expression under regulation by small RNAs}

\catchline{}{}{}{}{}

\title{Quantifying the noise in bursty gene expression under regulation by small RNAs}


\author{Shigang Qiu}

\address{College of Computer and Information Science, Southwest University, Chongqing, 400715, P. R. China}

\author{Tao Jia}

\address{College of Computer and Information Science, Southwest University, Chongqing, 400715, P. R. China\\
tjia@swu.edu.cn}

\maketitle

\begin{history}
\received{Day Month Year}
\revised{Day Month Year}
\end{history}

\begin{abstract}
Gene expression is a fundamental process in a living system. The small RNAs (sRNAs) is widely observed as a global regulator in gene expression. The inherent nonlinearity in this regulatory process together with the bursty production of  messenger RNA (mRNA), sRNA and protein make the exact solution for this stochastic process intractable. This is particularly the case when quantifying the protein noise level, which has great impact on multiple cellular processes. Here we propose an approximate yet reasonably accurate solution for the gene expression noise with infrequent burst and strong regulation by sRNAs. This analytical solution allows us to better analyze the noise and stochastic deviation of protein level. We find that the regulation amplifies the noise, reduces the protein level. The stochasticity in the regulation generates more proteins than what if the stochasticity is removed from the system. The sRNA level is most important to the relationship between the noise and stochastic deviation. The results provide analytical tools for more general studies of gene expression and strengthen our quantitative understandings of post-transcriptional regulation in controlling gene expression processes.

\end{abstract}

\ccode{PACS Nos.: 87.10.Mn, 02.50.r, 87.17.Aa}

\section{introduction}

Gene expression is inherently stochastic. The random nature of chemical reactions in a cell can make the variability inevitable in biological systems \cite{raser2005noise,elowitz2002stochastic,swain2002intrinsic,raser2004control,voulgarakis2017stochastic,bose2017allee}. Such intrinsic noise in gene expression can play crucial roles in critical cellular processes, even determine cell fates through certain cellular mechanisms, which are widely validated via experiments \cite{isaacs2003prediction,heitzler1991choice,eldar2010functional,kaern2005stochasticity,hu2011effects}. This noise can be amplified or compressed via different regulatory mechanisms during gene expression.
In eukaryotic cells, for example, the intrinsic noise can be modulated at the translational level, when regulator directly regulates the protein concentration, which plays a crucial role in several global regulatory networks \cite{blake2003noise,swain2002intrinsic,sanchez2013regulation,karmakar2016two}.
Likewise, post-transcriptional regulation is also a common regulatory mechanism \cite{gottesman2004small,storz2005abundance,shi2015post,bevilacqua2003post,thomson2006extensive}, in which regulators interact with mRNA to indirectly alter the protein synthesis. The post-transcriptional regulation process can be studied at a global level by analyzing the gene regulatory network to identify key regulators that are important in different regulatory pathways\cite{miller2016rsmw,kulkarni2014sequence,holmqvist2018rna}. At \cite{wang2018entangled} a more abstract level, we can also study quantitatively how the regulator can modify the gene expression level with given reaction scheme and parameters \cite{storz2011regulation,wagner2015chapter,jia2011intrinsic,wang2018entangled}, which is the focus of this paper.

Small non-coding RNAs (sRNAs) in bacteria are found as major players in post-transcriptional regulation \cite{wagner2015chapter,podkaminski2010small,mars2015small}. More than 80 sRNAs are cited for E.coli \cite{gottesman2011bacterial}, and the number is higher for other bacteria. The vast majority binds to mRNAs to modify the cell physiological function \cite{gottesman2005micros,podkaminski2010small}, so as to achieve its regulatory role. As an established mechanism, most of them act by base pairing with mRNAs followed by coupled stoichiometric degradation. Previous results have come out based on the corresponding model of coupled degradation \cite{platini2011regulation,levine2007quantitative}. However, the inherent nonlinearity of this reaction scheme makes the exact analytical solution intractable. Besides the nonlinearity of the problem, the difficulty in solving this problem also lies in the fact that the mRNAs created in each burst are not independent anymore: when one mRNA degrades with sRNA, the rest would face different regulatory concentration. Hence, proteins created by a burst of mRNAs can no longer be simply considered as a compound random variable of proteins produced by one single mRNA. This difficulty has been documented in previous works in which it is explicitly shown that the mean-field approach is not accurate in the limit of infrequent transcription events and strong sRNA-mRNA interactions \cite{platini2011regulation}.
While an approximate solution can be obtained \cite{kumar2016frequency}, it is only credible for a specific range of parameters.

In this paper, we aim to apply improved approximation methods to extend previous results. We analyzed the gene expression model based on infrequent bursting and sRNA regulation via strong interactions. We first give the approximate solution of the problem based on three assumptions that only valid for a limited range of parameters. We then release those assumptions and derived accurate expressions for mean protein levels and steady-state distributions that are valid for a wide range of parameters. Finally, we derive the noise and the stochastic deviation at the protein level, which serves as key features of this dynamical system. Our results extend previous works and provide insight into a quantitative understanding of the role of sRNA in post-transcriptional regulation.

\section{Model and Basic Results}

For the better reading, we first provide a table for the definition of parameters used in this paper. The definitions are also given in the main text of the paper when the parameter is first introduced.

\begin{table}[ht]
{\begin{tabular}{p{50pt}<{\centering}  p{280pt}} \toprule
Nomenclature & Description \\ \colrule
$k_m$ & mRNA synthesis rate \\
$\mu_m$ & mRNA degradation rate \\
$k_p$ & Protein production rate \\
$\mu_p$ & Protein degradation rate \\
$k_s$ & sRNA production rate \\
$\mu_s$ & sRNA degradation rate \\
$\gamma$ & Coupled degradation rate of mRNA and sRNA \\
$m$ & The number of mRNAs \\
$s$ & The number of sRNAs \\
$q_m$ & The degree of transcriptional burst size \\
$m_b$ & The number of mRNAs created in a transcriptional burst \\
$p_s$ & The number of proteins at the steady-state \\
$p_b$ &  The total number of proteins created in a burst \\
$p'_b$ & The number of proteins created from a single mRNA through a burst \\
$\tilde{p}_b$ & The total number of proteins created in a burst without regulator \\
$P_{mb}$ & mRNA burst size distribution \\
$P_{pb}$ & The total number of proteins burst size distribution \\
$P'_{pb}$ & The distribution of the number of proteins from a single mRNA burst \\
$\rho(s)$ & The steady-state distribution of sRNAs\\
$n_s$ & The number of sRNAs\\
$G_{pb}(z)$ & The generating function of the total number of proteins burst distribution \\
$G'_{pb}(z)$ & The generating function of the total number of proteins burst distribution from a single mRNA \\
$\tilde{G}_{pb}(z)$ & The generating function of the total number of protein burst distribution \\
$G_s(z,t)$ & The generating function of sRNA distribution at time t \\
\botrule
\end{tabular} \label{ta1}}
\end{table}

\begin{figure}[!t]
\centering
\includegraphics[width=8.5cm]{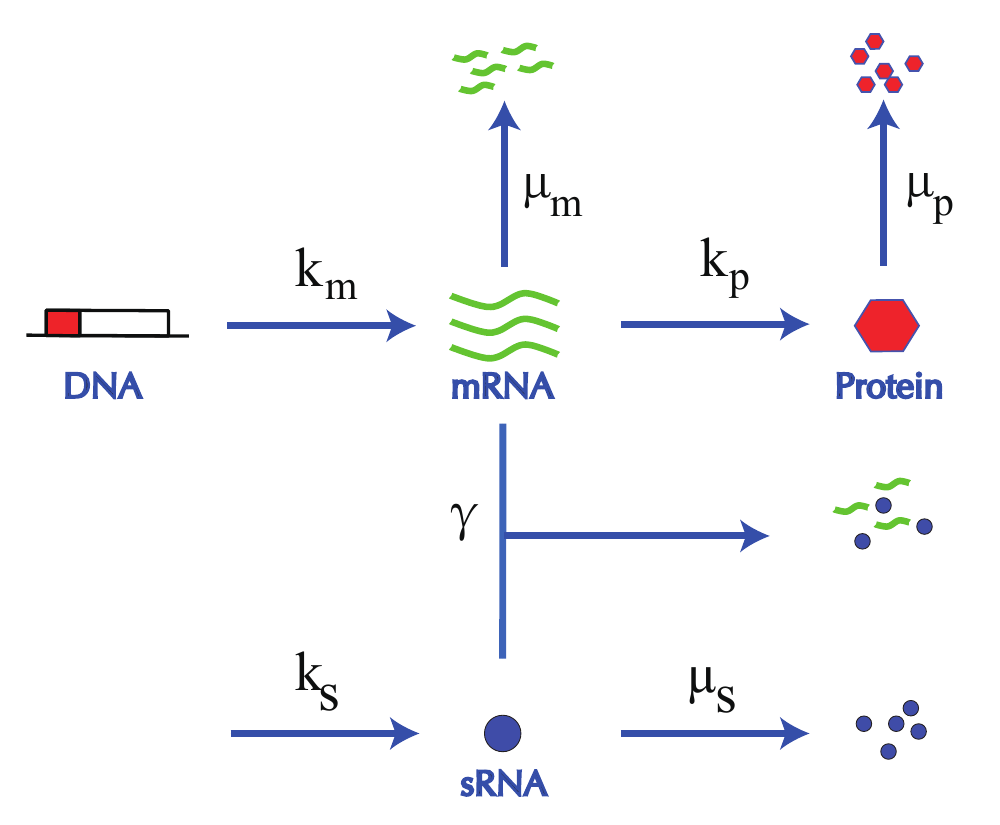}
\caption{The reaction scheme of post-transcriptional regulation by sRNA. The red hexagons denote protein, blue circles represent sRNA, and green lines show the bursty synthesis of mRNA. }
\label{1}
\end{figure}

The schematic diagram of gene regulatory network is shown in Fig. \ref{1}. We consider three elements in the system: mRNA, regulatory sRNA and protein. mRNA is assumed to be created in bursts, which occurs at rate $k_m$. Each burst of transcription produces a geometrically distributed number of mRNAs with the burst size distribution $P_{mb}(m)$ as \cite{jia2010post}
\begin{equation}
P_{mb}=(1-q_m)^{m-1}q_m,\label{eq1}
\end{equation}
in which the parameter $q_m$ controls the level of transcriptional burst with the mean burst size $\left\langle m_b\right\rangle=1/q_m$. Some experiments have also proved that it is a geometric distribution \cite{shahrezaei2008analytical,cai2006stochastic,yu2006probing}. The regulatory sRNA is assumed to be produced as a Poisson process at rate $k_s$ and degrade a Poisson process at rate $\mu_s$. In addition, as the regulator in gene expression, one sRNA can bind to a single mRNA and degrade together, which occurs at rate $\gamma$. In translation, the mRNA produces proteins with rate $k_p$ and the protein degrades with rate $\mu_p$.

The most common method to mathematically describe and solve this dynamical process is the master equation \cite{shahrezaei2008analytical,kumar2016frequency}. However, the existence of nonlinear terms makes the equation difficult to solve, even when only the mean and variance are considered. In the following, we will start with some extreme conditions under which the exact solution can be found. We then gradually release the constraints, move out from the extreme parameter regions that we start with and eventually come up with an approximate solution that is valid for a wide range of parameters.

To start with, we focus on the case that transcription rate $k_m$ is low and the mRNA degradation rate is much higher than the protein degradation rate ($\mu_m\gg\mu_p$), such that the bursts of proteins are typically well-separated in time and can be considered as independent events \cite{elgart2011connecting,elf2003fast,hayot2004linear}.
%
%
In particular, we use the following assumptions:
\begin{itemize}
\item[1)] mRNA degrades immediately on the appearance of sRNA;
\item[2)]There is no synthesis of new sRNAs \emph{during} a burst;
\item[3)]The distribution of sRNAs prior to a burst is the steady-state distribution of sRNAs in the absence of mRNAs, which is a Poisson distribution $\rho(s)=(n_s^s/s!)e^{-n_s}$, with the mean $n_s=\frac{k_s}{\mu_s}$.
\end{itemize}

For assumption 1), it implies that the regulation by sRNA is an instant modification of mRNA transcriptional burst level, which shows the strong sRNA-mRNA interactions. In other words, the number of mRNAs involved in translation depends on how much mRNA survived in the coupled degradation. Proteins are produced only when the number of mRNAs ($m$) is more than that of sRNAs ($s$) at the beginning of the burst. If so, $m-s$ mRNAs will proceed into translation process. Based on assumption 2), these $m-s$ mRNAs produce proteins as if there is no sRNAs regulation at all. We define $G_{pb}(z)=\sum_{n}z^n P_{pb}(n)$ as the generating function of the total number of proteins burst distribution, and $G'_{pb}(z)=\sum_{n}z^n P'_{pb}(n)$ as the generating function of protein burst distribution from a single mRNA. We have
\begin{equation}
G_{pb}(z)=\sum_{s,m}(P(m\leqslant s)+P(m>s)(G'_{pb}(z))^{m-s}).\label{eq2}
\end{equation}
Taking the distribution of random variable $s$ ($\rho(s)$) and $m$ ($P_{mb}(m)$) into this equation, we obtain
\begin{align}
G_{pb}(z)&=\sum_{i=j}^\infty \sum_{j=1}^\infty \rho(i)P_{mb}(j)+\sum_{i=0}^\infty \sum_{j=1}^\infty \rho(i)P_{mb}(i+j)G'^j_{pb}(z)\nonumber\\
&=1-\sum_{i=0}^\infty (1-q_m)^i\rho(i)\sum_{j=1}^\infty q_m(1-q_m)^{j-1}\nonumber\\
&+\sum_{i=0}^\infty (1-q_m)^i\rho(i)\sum_{j=1}^\infty q_m(1-q_m)^{j-1}G'^j_{pb}(z)\nonumber\\
&=1-e^{-n_s q_m}+e^{-n_s q_m}\frac{q_m G'_{pb}(z)}{1-G'_{pb}(z)(1-q_m)}.\label{eq3}
\end{align}
Eq. (\ref{eq3}) can be further simplified. In the absence of sRNA regulator, the generating function for the total number of protein burst distribution can be indicated as \cite{jia2010post}
\begin{equation}
\tilde{G}_{pb}(z)=\frac{q_m G'_{pb}}{1-G'_{pb}(1-q_m)},\label{eq4}
\end{equation}
where
\begin{equation}
G'_{pb}=\frac{\mu_m}{\mu_m+k_p(1-z)}.\nonumber
\end{equation}
Therefore, Eq. (\ref{eq3}) can be rewritten as \cite{kumar2016frequency}
\begin{equation}
G_{pb}(z)=1-e^{-n_s q_m}+e^{-n_s q_m}\tilde{G}_{pb}(z).\label{eq5}
\end{equation}
According to generating function, we can derive the mean, variance and squared coefficient of variance (noise strength). Denoting $p_b$ and $\tilde{p}_b$ as the random variable characterizing the number of proteins produced with and without regulation, respectively, we have
\begin{align}
\langle p_b\rangle &=e^{-n_s q_m}\langle \tilde{p}_b\rangle,\nonumber\\[3mm]
\sigma_{p_b}^2&=e^{-n_s q_m}\sigma_{\tilde{p}_b}^2+(e^{n_s q_m}-1)\langle p_b\rangle^2,\nonumber\\
\frac{\sigma_{p_b}^2}{\langle p_b\rangle^2}&=e^{n_s q_m}\frac{\sigma_{\tilde{p}_b}^2}{\langle \tilde{p}_b\rangle^2}+(e^{n_s q_m}-1),\label{eq6}
\end{align}
where the symbols $\langle .\rangle$ and $\sigma^2$ are used to denote the mean and variance. It is noteworthy Eq. (\ref{eq5}) and (\ref{eq6}) have the same terms, i.e., $e^{-n_s q_m}$. In Eq. (\ref{eq5}), this term is generated by the probability that at least one mRNA is left for translation. We can see that while the mean protein levels under regulation decreases by the factor $e^{-n_s q_m}$, the noise in protein burst size distribution increases. This is mainly due to the fact that regulation will give rise to a large probability of no or only very few proteins produced in the burst.

\section{Solution for a Wide Range of Parameters}

\begin{figure}
\centering
\includegraphics[width=13cm]{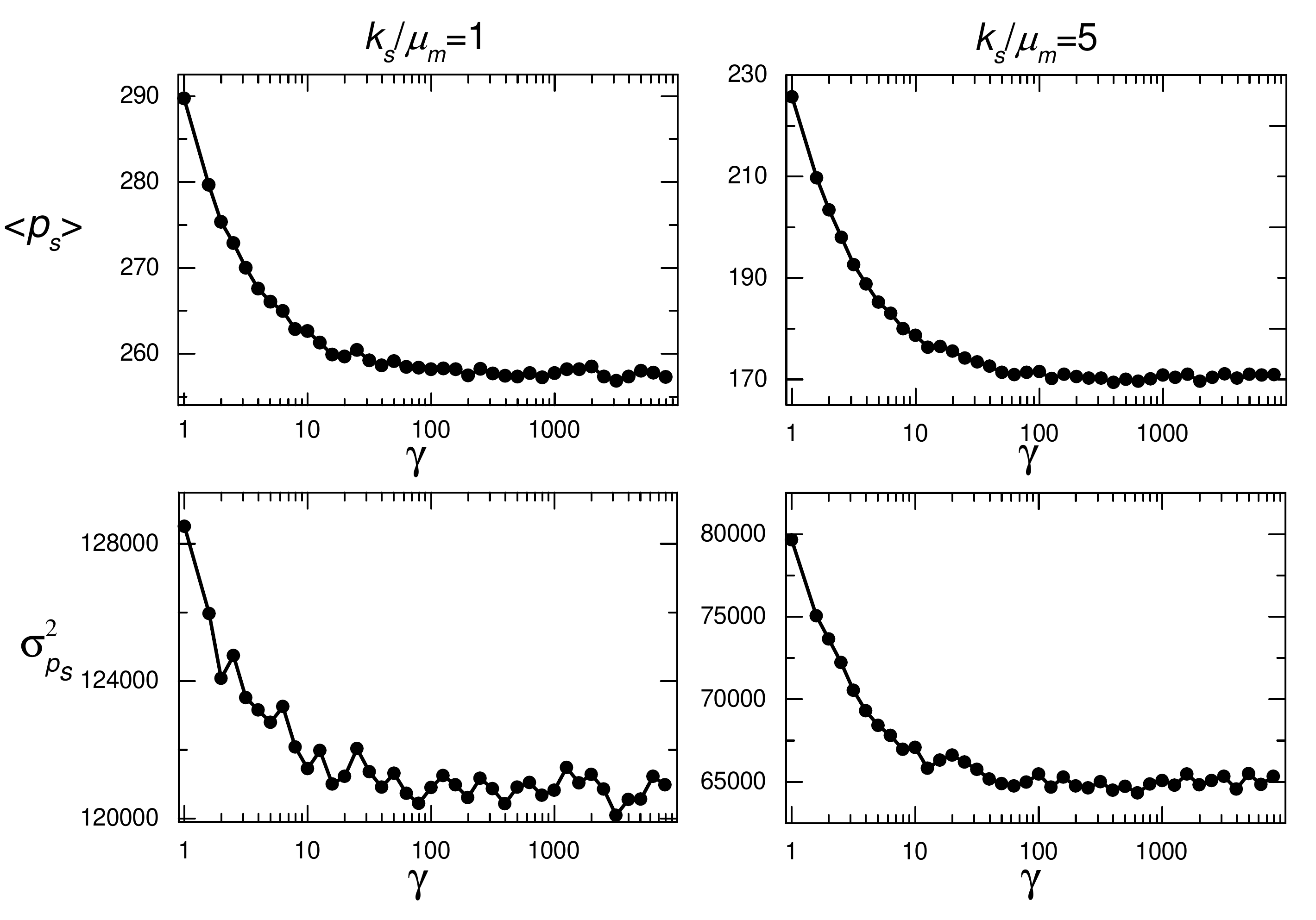}
\caption{The mean and variance of protein steady-state distributions based on simulation. In one case, $k_s=\mu_m$ and in the other case $\frac{k_s}{\mu_m}=5$. For both cases, $k_s=\mu_s=5$, $q_m=0.1$, $\frac{k_m}{\mu_m}=\frac{\mu_p}{\mu_m}=0.01$, $\frac{k_p}{\mu_m}=50$ and $\mu_m=1$. Both mean and variance become steady when $\gamma$ is large ($\gamma$ $>$ 10 max[$\mu_m, k_s$]).}
\label{2}
\end{figure}

The results obtained above are derived from three assumptions. Therefore, they are only valid for a limited range of parameters. To obtain results valid for a wider range of parameters, we need to find approximations that reasonably take the essential factors in the three assumptions. Let us first take assumption 1). The purpose of this assumption is to give the coupled degradation the most priority to occur among all reactions, as such that the reduction of sRNAs and mRNAs takes place instantaneously. This requires an infinitely large $\gamma$ value. However, since this assumption is to freeze the system dynamics during the coupled degradation, it can be approximately achieved when $\gamma$ value is much greater than $\mu_m$ and $k_s$. $\gamma\gg\mu_m$ is to make sure that the regulation is stronger than natural degradation and the number of proteins created during the coupled degradation can be neglected. $\gamma\gg k_s$ prevents the sRNA dynamics during the coupled degradation (the sRNA number is typically greater than 1 so $\gamma\gg\mu_s$ when $\gamma\gg k_s$). Based on simulation results shown in Fig. \ref{2}, we can see that when $\gamma \geqslant10$max[$k_s, \mu_m$], the mean and variance become stable and do not change very much as $\gamma$ further increases. Hence we can consider assumption 1) is approximately satisfied when $\gamma \geqslant10$max[$k_s, \mu_m$]\, which is also a feasible range experimentally \cite{mitarai2007efficient}.

Next, we focus on assumption 2), which implies that there is no sRNA creation during mRNA lifetime. The validity of this assumption is reflected by $\frac{k_s}{\mu_m}$ value, the mean number of sNRA that are created during the lifetime of an mRNA. To handle the case when $\frac{k_s}{\mu_m}$ is large, we need to re-derive the generating function of protein production process that is originally given by Eq. (\ref{eq3}).
Let us consider the case when mRNAs outnumber sRNAs and there are $m' = m - s$ mRNAs and 0 sRNA left after the initial mutual degradation. For the $m'$ mRNAs, we rank them by the \emph{inverse} order of their degradation. Specifically, the mRNA 1 is the last mRNA degraded whereas the mRNA $m'$ is the one that degrades first. Denote $T_i$ as the time interval between the degradation of mRNA $i$ and mRNA $i+1$ and $T_{m'}$ is the time elapsed until the degradation of the first mRNA. As the degradation of mRNAs is considered to be a Poisson process, the waiting time $T_i$ follows the exponential distribution with mean $1/i\mu_m$ without regulation. Based on assumption 1), the creation of one sRNA will result in the immediate degradation of one mRNA. Hence, when regulation is taken into account, $T_i$ still follows with exponential distribution but the mean will be $1/(i\mu_m+k_s)$.

We Define $G_i(z)$ as the generating function of proteins created during a time period
$T_i$. Based on results introduced previously \cite{jia2010post}, we have
\begin{equation}
G_i(z)=\frac{i\mu_m+k_s}{i\mu_m+k_s+ik_p(1-z)}.\label{eq7}
\end{equation}
Because the time interval $T_i$'s are independent, we can multiply the $G_i(z)$'s together to get the generating function of the total number of protein burst distribution (i.e. the $G'_{pb}$ that mentioned above). Note that the overall protein creation rate during $T_i$ is $ik_p$. Therefore, $G'_{pb}$ conditioned on $m'$ mRNA left can be modified to
\begin{equation}
G'_{pb}(z)=\prod_{i=1}^{m'} \frac{i\mu_m+k_s}{i\mu_m+k_s+ik_p(1-z)}.\label{eq8}
\end{equation}
By taking the probability that $m'$ mRNAs survive into account, we have
\begin{align}
G_{pb}(z)&=\sum_{i=j}^\infty \sum_{j=1}^\infty \rho(i)P_{mb}(j)\nonumber\\
&+\sum_{i=0}^\infty \sum_{j=1}^\infty \rho(i)P_{mb}(i+j)\prod_{i=1}^j \frac{i\mu_m+k_s}{i\mu_m+k_s+ik_p(1-z)}\nonumber\\
&=1-e^{-n_s q_m}+e^{-n_s q_m}\nonumber\\
&\times \sum_{j=1}^\infty q_m(1-q_m)^{j-1}\prod_{i=1}^j \frac{i\mu_m+k_s}{i\mu_m+k_s+ik_p(1-z)}.\label{eq9}
\end{align}
Unfortunately, we can not simplify Eq.(\ref{eq9}) further. From the generating function, we can derive the mean protein burst size as
\begin{equation}
\langle p_b\rangle=e^{-n_s q_m}\sum_{j=1}^\infty q_m(1-q_m)^{j-1}\sum_{i=1}^j \frac{ik_p}{i\mu_m+k_s}.\label{eq10}
\end{equation}
The last term in Eq. (\ref{eq10}) can be further simplified as
\begin{align}
\sum_{i=1}^j \frac{ik_p}{i\mu_m+k_s}&=\frac{k_p}{\mu_m}\sum_{i=1}^j \left(1-\frac{k_s}{i\mu_m+k_s}\right)\nonumber\\
&=\frac{k_p}{\mu_m}\left(j-\frac{k_s}{\mu_m}\sum_{i=1}^j\frac{1}{i+\frac{k_s}{\mu_m}}\right)\nonumber\\
&=\frac{k_p}{\mu_m}\left(j+\frac{k_s}{\mu_m}\ln\frac{k_s}{k_s+j\mu_m}\right),\label{eq11}
\end{align}
where the approximation of harmonic number is applied. The harmonic number is defined as
\begin{equation}
H(n)=\sum_{k=1}^n \frac{1}{k}\cong\ln(n)+\gamma,\nonumber
\end{equation}
where $\gamma$ is the Euler-Mascheroni constant. Furthermore, we have
\begin{align}
\sum_{k=1}^n \frac{1}{k+C}&=H(C+n)-H(C)\nonumber\\
&=\ln\left(\frac{C+n}{C}\right)\nonumber
\end{align}
where $C$ is a constant. Taking Eq. (\ref{eq11}) into Eq. (\ref{eq10}), we obtain
\begin{align}
\langle p_b\rangle=&e^{-n_s q_m}\frac{k_p}{\mu_m}\sum_{j=1}^\infty q_m(1-q_m)^{j-1}\left(j+\frac{k_s}{\mu_m}\ln\frac{k_s}{k_s+j\mu_m}\right)\nonumber\\
=&e^{-n_s q_m}\frac{k_p}{\mu_m}\Bigg(\frac{1}{q_m}+\frac{k_s}{\mu_m}\bigg(\ln k_s\nonumber\\
&-\sum_{j=1}^\infty q_m(1-q_m)^{j-1}\ln(k_s+j\mu_m)\bigg)\Bigg)\nonumber\\
=&e^{-n_s q_m}\frac{k_p}{\mu_m}\Bigg(\frac{1}{q_m}+\frac{k_s}{\mu_m}\ln\frac{k_s}{\mu_m}+\frac{k_s q_m}{\mu_m}\nonumber\\
&\times\Phi^{(0,1,0)}\left(1-q_m,0,\frac{k_s}{\mu_m}+1\right)\Bigg)\label{eq12}
\end{align}
where $\Phi$ is the Hurwitz-Lerch transcendental function defined as
\begin{equation}
\Phi(z,s,a)=\sum_{k=0}^\infty z^k(k+a)^{-s},\nonumber
\end{equation}
and $\Phi^{(0,1,0)}(z,s,a)$ indicate the partial derivatives respect to $s$.

Finally, we come to assumption 3) which says that the number of sRNAs prior to a transcription event follows Poisson distribution with mean $n_s=k_s/\mu_s$. Without the appearance of mRNA, sRNA evolves according to the standard birth and death process which gives a Poisson distribution in steady-state. The assumption 3) holds when each transcription occurs very infrequently (very small $k_m$ value). However, when $k_m$ is not that small, we have to consider the transient behavior of the sRNA evolution.
\begin{figure}[!t]
\centering
\includegraphics[width=13cm]{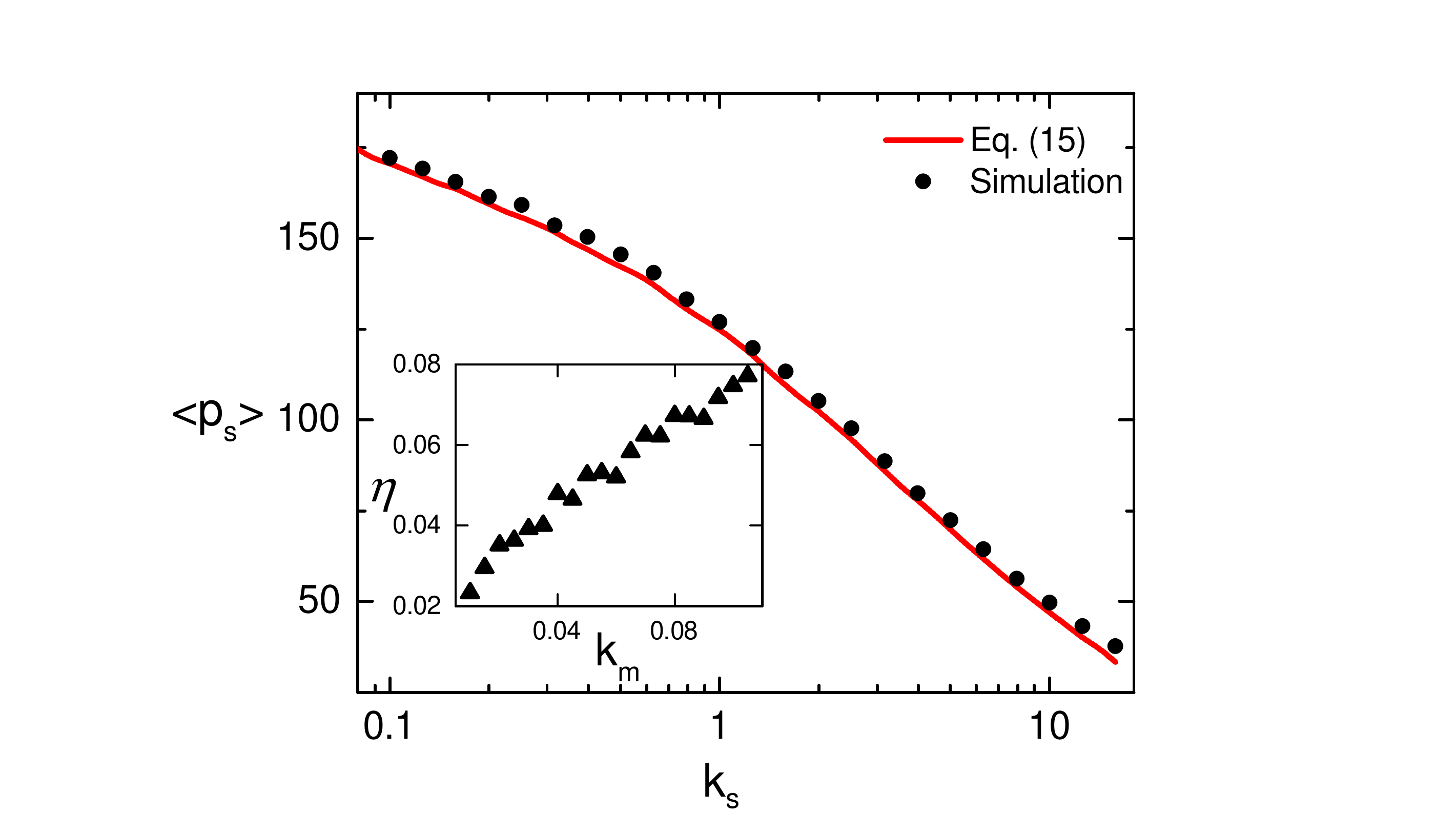}
\caption{The steady-state mean protein number vs $k_s$. The calculation based on Eq. (\ref{eq15}) is very close to simulation result. The parameters are chosen as $\mu_m=1$, $\frac{k_s}{\mu_s}=2$, $k_m=\mu_p=0.01$, $q_m=0.2$, $\gamma=100$ and $k_p=50$. In the insert, we show the relative error $\eta$ vs $k_m$. The error increases as $k_m$ increases. The parameters used in the inset are $\mu_m=\mu_s=1$, $k_s=2$, $\mu_p=0.01$, $q_m=0.2$, $\gamma=50$ and $k_p=50$.}
\label{3}
\end{figure}

Define $G_s(z,t)$ as the generating function of sRNA distribution at time t. $t=0$ is the time that the previous translational burst ends, i.e., all mRNAs created in one transcriptional burst are degraded. In the problem analyzed, we are more interested in situations that the regulation tunes the protein level but does not fully repress the translation. This corresponds to the parameter region that $1/q_m\gg n_s$. In this parameter region, the mRNAs produced in transcriptional burst usually outnumber the sRNAs and typically there is no sRNA left after the regulation. This provides the initial condition that the number of sRNAs is zero at $t=0$. Based on the results of birth and death process \cite{medhi2002stochastic}, we have
\begin{equation}
G_s(z,t)=exp[-\frac{k_s}{\mu_s}(1-e^{-\mu_s t})(1-z)].\label{eq13}
\end{equation}
The waiting time distribution that the next burst occurs is exponential with mean $1/k_m$. Then the generating function of sRNA distribution prior to the burst (distribution $\rho(s)$) will be
\begin{equation}
G_s(z)=\int_0^\infty exp[-\frac{k_s}{\mu_s}(1-e^{-\mu_s t})(1-z)]\times k_m e^{k_m t}dt.\label{eq14}
\end{equation}

Note that in the above deviation, the term that contributes to the final result is $\sum_{i=0}^\infty(1-q_m)^i\rho(i)$ (see Eq. (\ref{eq3})). By recalling the definition of generating function $G_s(z)=\sum_{i=0}^\infty z^i\rho(i)$, we notice that the term $\sum_{i=0}^\infty(1-q_m)^i\rho(i)$ equals $G_s(1-q_m)$. Thus a more accurate expression of the results presented (Eq. (\ref{eq9})- Eq. (\ref{eq12})) is to replace the term $e^{-n_s q_m}$ by $G_s(1-q_m)$. On the other hand, the form of $G_s(1-q_m)$ is complicated. Based on numerical evaluation, we can approximate $\rho(i)$ by a Poisson distribution with the mean given by $G_s(z)$. This only requires replacing the term $n_s=k_s/\mu_s$ in $e^{-n_s q_m}$ by $n'_s=k_s/(\mu_s+k_m)$, which given a more simple form and very close to $G_s(1-q_m)$.

Given the protein burst size distribution, we can connect it to the protein
steady-state level \cite{elgart2011connecting}, which gives
\begin{align}
\langle p_s\rangle=&\frac{k_m k_p}{\mu_m\mu_p}e^{-\frac{k_s q_m}{\mu_s+k_m}}\bigg(\frac{1}{q_m}+\frac{k_s}{\mu_m}\ln\frac{k_s}{\mu_m}+\frac{k_s q_m}{\mu_m}\nonumber\\
&\times\Phi^{(0,1,0)}\Big(1-q_m,0,\frac{k_s}{\mu_m}+1\Big)\bigg).\label{eq15}
\end{align}

The result in Eq. (\ref{eq15}) is tested by simulation for a range of parameters with $\frac{k_s}{\mu_m}\in [0.1, 10]$ and $\frac{k_m}{\mu_m}\in [0.01, 0.1]$. The plot (Fig. \ref{3}) shows a perfect match between Eq. (\ref{eq15}) and the simulation when $k_m$ is small. While the error does not depend on $k_s$, it increases when $k_m$ becomes large. This is because our deduction is based on the infrequent transcription that the bursts are clearly separated in time. Increasing $k_m$ will violate this assumption and bring more error. However, even take this into consideration, Eq.(\ref{eq15}) still accurately quantifies the steady-state mean protein levels (within 8\% error) for the range of parameters tested.

\section{Noise regulation and the Stochastic Deviation}

The deduction above provides us a better analytical tool to analyze the impact of post-transcriptional regulation. Here we focus on two quantities. The first the noise level in protein steady-state distribution that is usually characterized by the squared coefficient of variance. The second is the stochastic deviation of protein steady-state level, which is quantified as the difference of mean protein number between the stochastic and deterministic system \cite{kuwahara2012stochastic,samoilov2006deviant}.

\begin{figure}
\centering
\includegraphics[width=13cm]{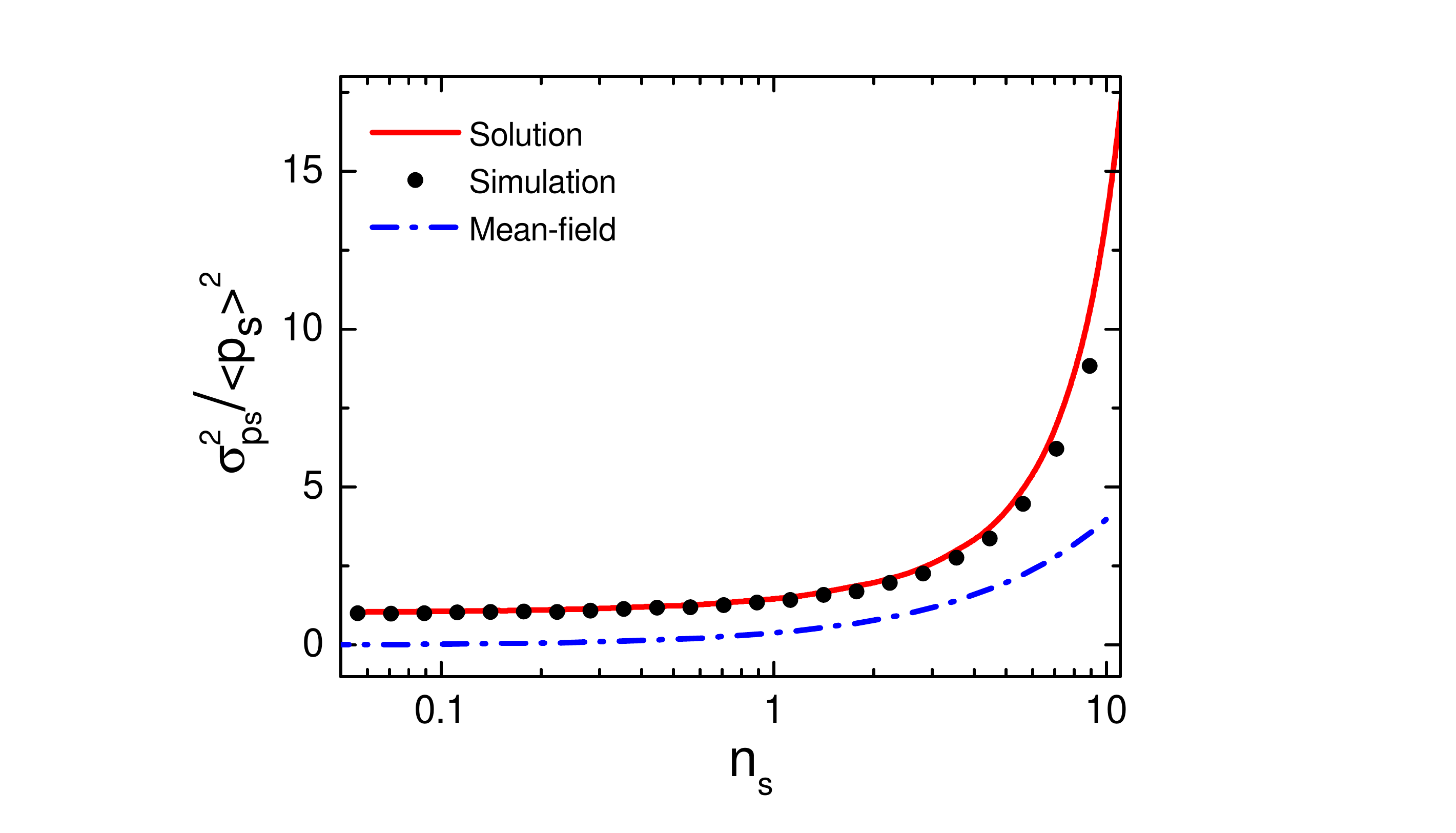}
\caption{The noise in protein steady-state distribution as a function of the mean sRNA levels $n_s$. The analytical result (the red line) calculated based on Eq. (\ref{eq16}) is very close to stochastic simulations for a range of parameters, while the mean-field result (the blue dashed line) always underestimates the noise level. The parameters are $\mu_m=\mu_s=1$, $k_m=\mu_p=0.01$, $q_m=0.2$, $\gamma=100$ and $k_p=50$. The noise for mean-field case is calculated using the assumption of mean-field approach that the protein steady-state distribution follows a Poisson distribution. Hence the squared coefficient of variance equals $1/\langle p_s\rangle $}
\label{4}
\end{figure}

The squared coefficient of variance (noise level) of protein steady-state distribution can be found using the protein burst size distribution \cite{elgart2011connecting}, as
\begin{equation}
\frac{\sigma^2_{p_s}}{\langle p_s\rangle^2}=\frac{1}{\langle p_s\rangle}+\frac{\mu_p}{2k_m}(1+\frac{\sigma^2_{p_b}}{\langle p_b\rangle^2}-\frac{1}{\langle p_b\rangle}).\label{eq16}
\end{equation}
The $\langle p_s\rangle$ is readily given in Eq.(\ref{eq15}) and the mean and variance of $p_b$ can be derived from the generating function in Eq.(\ref{eq9}). The analytical calculation introduced in this paper matches the simulation well (Fig. \ref{4}). The noise increases when the mean sRNA number increases, indicating that the system is likely to have more fluctuations when the regulation is activated. The actual noise level is always greater than what mean-field predicts \cite{platini2011regulation}, demonstrating that our approach is more appropriate for this stochastic and nonlinear system.

Although the mean-field method is widely used \cite{platini2011regulation,mendoncca2016inactive,dhiman2018steady}, it cannot accurately describe the stochastic system. we further move to another quantity to quantify the discrepancy due to stochasticity. The stochastic deviation is described as the deviation between stochastic and deterministic system \cite{samoilov2006deviant}. Previous studies on the gene expression process found that the mean mRNA levels predicted by the stochastic model are larger than that predicted by the deterministic model \cite{kuwahara2012stochastic}, demonstrating the importance of the stochastic approach to gene expression process. Here we follow a similar approach to analyze the protein level when regulation and mRNA and protein bursts are involved.

\begin{figure}
\centering
\includegraphics[width=13cm]{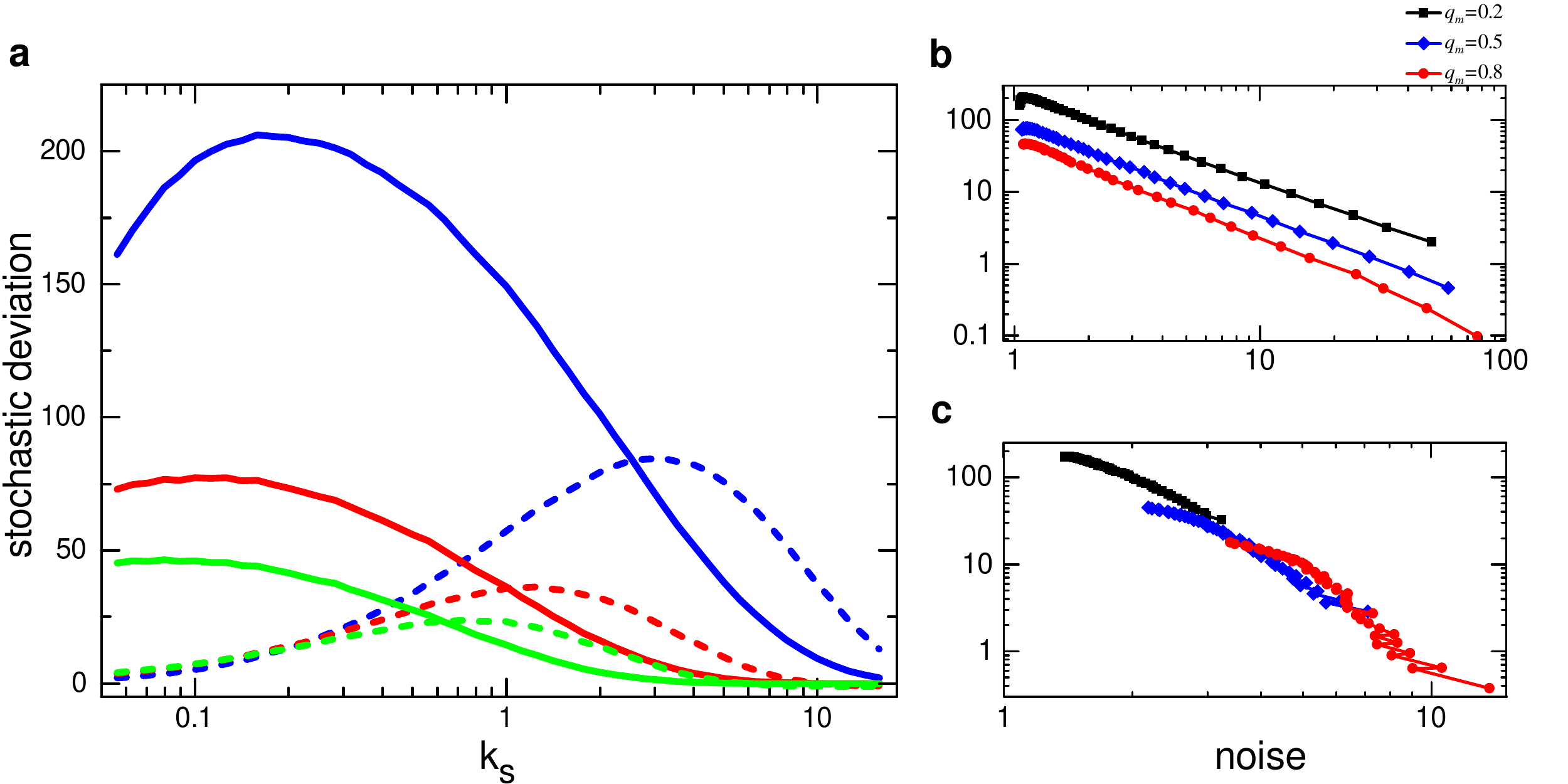}
\caption{Stochastic deviation in post-transcriptional regulation. The mean levels of protein in deterministic model is derived from the mean-field method and the the mean levels of protein in stochastic model is based on Eq. (\ref{eq15}). (a) Variations of stochastic deviation plotted as a function of $k_s$ for the case $\mu_s=1$ (solid line) and $k_s/\mu_s=2$ (dashed line) for three different values of $q_m$: 0.2 (blue), 0.5 (red), and 0.8 (green). (b) Variation of stochastic deviation with noise for $\mu_s=1$ for three values of $q_m$. (c) Variations of stochastic deviation  in the case that the mean sRNA levels unchanged $k_s/\mu_s=2$, with noise for three values of $q_m$. In all plots, other parameters are: $k_m=0.01$, $\mu_m=1$, $k_p=50$, $\mu_p=0.01$, $\gamma=100$.}
\label{5}
\end{figure}

As shown in Fig. \ref{5}a, the stochastic deviation changes non-monotonically with the sRNA level. The peak roughly appears at the point when the sRNA level and mRNA burst size are comparable. This indicates that the stochastic deviation is large when the gene expression is under regulation. While the regulation is to decrease the protein level, the stochasticity actually helps the cell to gain more proteins.

Stochastic deviation and noise are two different types of stochastic effect. Both of them can be affected by sNRA and mRNA dynamics. Indeed, previous work has shown that the noise and nonlinear of the system are the main sources of stochastic deviation \cite{samoilov2006deviant,shi2015post}. But here the noise level can be tuned by either sRNA or mRNA. Hence, it is unclear which one is most relevant to the stochastic deviation. For this reason, we analyze the relationship between stochastic deviation and noise, when the sRNA and mRNA level are controlled (Fig. \ref{5}b, c). We find that when the sRNA level is fixed, different mRNA levels give almost the same curve for the change of stochastic deviation with noise. In other words, when the sRNA level is given, the gain of proteins due to stochasticity will not change dramatically for the same value of noise, even mean mRNA burst size is different. Thus, under the post-transcriptional regulation, the stochastic deviation will decrease as noise get large, and it is only related to the sRNA synthesis rate, but the mean level of mRNA.

\section{Conclusion}
In summary, we studied the gene expression process when mRNA and protein bursts and sRNA regulation are involved. We extend previous works by proposing an approximate yet reasonably good solution to the protein steady-state distribution. The approach not only gives a useful tool to analytically study the stochastic system this paper currently focuses on, but may also motivate other methods to handle similar yet more complicated system, i.e., involved genetic switches \cite{spudich1976non,korobkova2006hidden,roy2001cooperative} or feedback regulation \cite{swain2004efficient,kumar2014exact,liang2017universal}. Using the analytical results obtained, we further analyze the noise and the stochastic deviation of protein steady-state level. We find that the regulation amplifies the noise, reduces the protein level. But on the other hand, the stochasticity in the regulation generates more proteins than if the stochasticity is removed from the system. We also find a scaling-like phenomenon in the relationship between the stochastic deviation and the noise. Once the sRNA level is fixed, the gain of proteins due to stochasticity will not change dramatically for the same value of noise, even mean mRNA burst size is different.

Our work provides insight into how different mechanisms of post-transcriptional regulation can be used to fine-tune the noise and the stochastic deviation in gene expression with potential implications for studies addressing the evolutionary importance of noise and deviation in biological systems. The analytical results show an effective method for accurate quantitative modeling of stochastic cellular processes in large regions of parameter space. However, it is noteworthy that it is unclear if our research display all dynamic relationships of the motif considered in this work. It is hoped that our findings will be experimentally validated in the future to provide a quantitative understanding of the role of post-transcriptional regulation in gene expression.

\section{Acknowledgements}

This work is supported by the Natural Science Foundation of China (No. 61603309).



\begin{thebibliography}{10}

\bibitem{raser2005noise}
J.~M. Raser and E.~K. O'shea, {\em Science} {\bf 309}, 2010  (2005).

\bibitem{elowitz2002stochastic}
M.~B. Elowitz, A.~J. Levine, E.~D. Siggia and P.~S. Swain, {\em Science} {\bf
  297}, 1183  (2002).

\bibitem{swain2002intrinsic}
P.~S. Swain, M.~B. Elowitz and E.~D. Siggia, {\em Proc. Natl. Acad. Sci.} {\bf
  99}, 12795  (2002).

\bibitem{raser2004control}
J.~M. Raser and E.~K. O'shea, {\em science} {\bf 304}, 1811  (2004).

\bibitem{voulgarakis2017stochastic}
N.~K. Voulgarakis, {\em Int. J. Mod. Phys. C} {\bf 28}, 1750102  (2017).

\bibitem{bose2017allee}
I.~Bose, M.~Pal and C.~Karmakar, {\em Int. J. Mod. Phys. C} {\bf 28}, 1750074
  (2017).

\bibitem{isaacs2003prediction}
F.~J. Isaacs, J.~Hasty, C.~R. Cantor and J.~J. Collins, {\em Proc. Natl. Acad.
  Sci.} {\bf 100}, 7714  (2003).

\bibitem{heitzler1991choice}
P.~Heitzler and P.~Simpson, {\em Cell} {\bf 64}, 1083  (1991).

\bibitem{eldar2010functional}
A.~Eldar and M.~B. Elowitz, {\em Nature} {\bf 467}, 167  (2010).

\bibitem{kaern2005stochasticity}
M.~K{\ae}rn, T.~C. Elston, W.~J. Blake and J.~J. Collins, {\em Nat. Rev.
  Genet.} {\bf 6}, 451  (2005).

\bibitem{hu2011effects}
B.~Hu, D.~A. Kessler, W.-J. Rappel and H.~Levine, {\em Phys. Rev. Lett.} {\bf
  107}, 148101  (2011).

\bibitem{blake2003noise}
W.~J. Blake, M.~K{\ae}rn, C.~R. Cantor and J.~J. Collins, {\em Nature} {\bf
  422}, 633  (2003).

\bibitem{sanchez2013regulation}
A.~Sanchez, S.~Choubey and J.~Kondev, {\em Annu. Rev. Biophys.} {\bf 42}, 469
  (2013).

\bibitem{karmakar2016two}
R.~Karmakar, {\em Int. J. Mod. Phys. C} {\bf 27}, 1650056  (2016).

\bibitem{gottesman2004small}
S.~Gottesman, {\em Annu. Rev. Microbiol.} {\bf 58}, 303  (2004).

\bibitem{storz2005abundance}
G.~Storz, S.~Altuvia and K.~M. Wassarman, {\em Annu. Rev. Biochem.} {\bf 74},
  199  (2005).

\bibitem{shi2015post}
C.~Shi, S.~Wang, T.~Zhou and Y.~Jiang, {\em Phys. Biol.} {\bf 12}, 056002
  (2015).

\bibitem{bevilacqua2003post}
A.~Bevilacqua, M.~C. Ceriani, S.~Capaccioli and A.~Nicolin, {\em J. Cell.
  Physiol.} {\bf 195}, 356  (2003).

\bibitem{thomson2006extensive}
J.~M. Thomson, M.~Newman, J.~S. Parker, E.~M. Morin-Kensicki, T.~Wright and
  S.~M. Hammond, {\em Genes Dev.} {\bf 20}, 2202  (2006).

\bibitem{miller2016rsmw}
C.~L. Miller, M.~Romero, S.~R. Karna, T.~Chen, S.~Heeb and K.~P. Leung, {\em
  BMC Microbiol.} {\bf 16}, p. 155  (2016).

\bibitem{kulkarni2014sequence}
P.~R. Kulkarni, T.~Jia, S.~A. Kuehne, T.~M. Kerkering, E.~R. Morris, M.~S.
  Searle, S.~Heeb, J.~Rao and R.~V. Kulkarni, {\em Nucleic Acids Res.} {\bf
  42}, 6811  (2014).

\bibitem{holmqvist2018rna}
E.~Holmqvist and J.~Vogel, {\em Nat. Rev. Microbiol.} , p.~1  (2018).

\bibitem{wang2018entangled}
H.~Wang, P.~Liu, Q.~Li and T.~Zhou, {\em FEBS Lett.} {\bf 592}, 1135  (2018).

\bibitem{storz2011regulation}
G.~Storz, J.~Vogel and K.~M. Wassarman, {\em Mol. Cell} {\bf 43}, 880  (2011).

\bibitem{wagner2015chapter}
E.~G.~H. Wagner and P.~Romby, {\em Adv. Genet.} {\bf 90}, 133  (2015).

\bibitem{jia2011intrinsic}
T.~Jia and R.~V. Kulkarni, {\em Phys. Rev. Lett.} {\bf 106}, p. 058102  (2011).

\bibitem{podkaminski2010small}
D.~Podkaminski and J.~Vogel, {\em Mol. Microbiol.} {\bf 78}, 1327  (2010).

\bibitem{mars2015small}
R.~A. Mars, P.~Nicolas, M.~Ciccolini, E.~Reilman, A.~Reder, M.~Schaffer,
  U.~M{\"a}der, U.~V{\"o}lker, J.~M. van Dijl and E.~L. Denham, {\em PLoS
  Genet.} {\bf 11}, e1005046  (2015).

\bibitem{gottesman2011bacterial}
S.~Gottesman and G.~Storz, {\em Cold Spring Harbor Perspect. Biol.} {\bf 3},
  a003798  (2011).

\bibitem{gottesman2005micros}
S.~Gottesman, {\em Trends Genet.} {\bf 21}, 399  (2005).

\bibitem{platini2011regulation}
T.~Platini, T.~Jia and R.~V. Kulkarni, {\em Phys. Rev. E} {\bf 84}, 021928
  (2011).

\bibitem{levine2007quantitative}
E.~Levine, Z.~Zhang, T.~Kuhlman and T.~Hwa, {\em PLoS. Biol.} {\bf 5}, e229
  (2007).

\bibitem{kumar2016frequency}
N.~Kumar, T.~Jia, K.~Zarringhalam and R.~V. Kulkarni, {\em Phys. Rev. E} {\bf
  94}, 042419  (2016).

\bibitem{jia2010post}
T.~Jia and R.~V. Kulkarni, {\em Phys. Rev. Lett.} {\bf 105}, 018101  (2010).

\bibitem{shahrezaei2008analytical}
V.~Shahrezaei and P.~S. Swain, {\em Proc. Natl. Acad. Sci.} {\bf 105}, 17256
  (2008).

\bibitem{cai2006stochastic}
L.~Cai, N.~Friedman and X.~S. Xie, {\em Nature} {\bf 440}, 358  (2006).

\bibitem{yu2006probing}
J.~Yu, J.~Xiao, X.~Ren, K.~Lao and X.~S. Xie, {\em Science} {\bf 311}, 1600
  (2006).

\bibitem{elgart2011connecting}
V.~Elgart, T.~Jia, A.~T. Fenley and R.~Kulkarni, {\em Phys. Biol.} {\bf 8},
  046001  (2011).

\bibitem{elf2003fast}
J.~Elf and M.~Ehrenberg, {\em Genome Res.} {\bf 13}, 2475  (2003).

\bibitem{hayot2004linear}
F.~Hayot and C.~Jayaprakash, {\em Phys. Biol.} {\bf 1}, 205  (2004).

\bibitem{mitarai2007efficient}
N.~Mitarai, A.~M. Andersson, S.~Krishna, S.~Semsey and K.~Sneppen, {\em Phys.
  Biol.} {\bf 4}, 164  (2007).

\bibitem{medhi2002stochastic}
J.~Medhi, {\em Stochastic models in queueing theory} (Academic Press, 2002).

\bibitem{kuwahara2012stochastic}
H.~Kuwahara and R.~Schwartz, {\em J. R. Soc. Interface} , rsif20110757  (2012).

\bibitem{samoilov2006deviant}
M.~S. Samoilov and A.~P. Arkin, {\em Nat. Biotechnol.} {\bf 24}, 1235  (2006).

\bibitem{mendoncca2016inactive}
J.~R.~G. Mendon{\c{c}}a, {\em Int. J. Mod. Phys. C} {\bf 27}, 1650016  (2016).

\bibitem{dhiman2018steady}
I.~Dhiman and A.~K. Gupta, {\em Int. J. Mod. Phys. C} {\bf 29}, 1850037
  (2018).

\bibitem{spudich1976non}
J.~L. Spudich and D.~E. Koshland, {\em Nature} {\bf 262}, 467  (1976).

\bibitem{korobkova2006hidden}
E.~A. Korobkova, T.~Emonet, H.~Park and P.~Cluzel, {\em Phys. Rev. Lett.} {\bf
  96}, 058105  (2006).

\bibitem{roy2001cooperative}
S.~Roy, I.~Bose and S.~S. Manna, {\em Int. J. Mod. Phys. C} {\bf 12}, 413
  (2001).

\bibitem{swain2004efficient}
P.~S. Swain, {\em J. Mol. Biol.} {\bf 344}, 965  (2004).

\bibitem{kumar2014exact}
N.~Kumar, T.~Platini and R.~V. Kulkarni, {\em Phys. Rev. Lett.} {\bf 113},
  268105  (2014).

\bibitem{liang2017universal}
J.~Liang, Y.~Hu, G.~Chen and T.~Zhou, {\em Sci. Rep.} {\bf 7}, p. 42857
  (2017).

\end{thebibliography}
\end{document}